\documentclass[twocolumn,preprintnumbers]{revtex4}
\usepackage{bm,color}
\usepackage{graphicx}
\usepackage{amsmath}
\usepackage{ulem}
\usepackage{CJKutf8,CJK,CJKspace,CJKpunct}

\begin{document}
\title{Bulk and surface topological indices for a skyrmion string}
\author{Wataru Koshibae$^{1}$}
\email{wataru@riken.jp}
\author{Naoto Nagaosa$^{1,2}$}
\email{nagaosa@riken.jp}
\affiliation{$^1$ RIKEN Center for Emergent Matter Science (CEMS), 
Wako, Saitama 351-0198, Japan}
\affiliation{$^2$ Department of Applied Physics, The University of Tokyo, 7-3-1, 
Hongo, Bunkyo-ku, Tokyo 113-8656, Japan}

\begin{abstract}
The magnetic skyrmion is a topological magnetic vortex, and its topological nature is characterized by an index called skyrmion number which is a mapping of the magnetic moments defined on a two-dimensional space to a unit sphere. 
In three-dimensions, a skyrmion, i.e., a vortex penetrating though the magnet 
naturally forms a string, which terminates at the surfaces of the magnet or in the bulk. 
For such a string, the topological indices, which control its topological stability are less trivial. 
Here, we show theoretically, in terms of numerical simulation for the current-driven motion of a skyrmion string in a film sample with the step edges on the surface, that the topological indices relevant to the stability are the followings; (i) skyrmion number along the developed surface, and (ii) the monopole charge in the bulk defined as the integral over the surface enclosing a singular magnetic configuration. As long as the magnetic configuration is slowly varying, the former is conserved while its changes is associated with nonzero monopole charge. The skyrmion number and the monoplole charge offer a coherent understanding of the stability of the topological magnetic texture and the nontrivial dynamics of skyrmion strings.  
\end{abstract}

\maketitle
%\newpage

Magnetic skyrmion, a swirling magnetic vortex has attracted much attention in recent years~\cite{Bogdanov1,Bogdanov2,Rosler,Binz,Tewari,
Muhlbauer09,Munzer10,YuXZ10N,YuXZNM11,NagTok}. 
The main focus is on its topological  nature: 
the skyrmion is topologically distinguished from ferromagnetic state for instance, 
i.e., these magnetic textures cannot be related to each other 
within continuous deformation.  
This topological difference is characterized by 
the skyrmion number $N_{\rm sk}$.  
To make the definition of the index $N_{\rm sk}$ clear, 
for given normalized magnetic moments  
$\{\bm n_{\bm r}\}_{\bm r\in\Lambda}$ on the set of lattice sites $\Lambda$, 
we define  
\begin{align}
N_{\rm topol}(\Omega)=&\frac{1}{2\pi}\int_{\Omega} b_\text{normal} d\omega, 
\label{eqn:topo}
\end{align}
where $b_\text{normal}=\bm b\cdot\bm e$ with the emergent 
$b$-field $b_i=(1/4)\varepsilon_{ijk}\bm n\cdot(\partial_j\bm n\times\partial_k\bm n)$~\cite{Zang,Schulz,NT} and 
$\bm e$ is the normal unit vector to the two-dimensional domain of integral $\Omega\subset\Lambda$. 
(This $N_{\rm topol}$ is a functional of $\{\bm n_{\bm r}\}_{\bm r\in\Omega}$ and 
depends on time for the dynamics, but we will not explicitly write 
those degrees of freedom in the expression Eq.~(\ref{eqn:topo}).) 
Usually, the skyrmion number is defined as 
$N_{\rm sk,\Omega}=N_{\rm topol}(\Omega)$ where 
$\Omega$ is a plane perpendicular to the external magnetic field and the direction 
${\bm e}$ is taken to be parallel to the magnetic field. 
Under the condition where $\bm n_{\bm r}\to{\bm e}$ for $|\bm r|\to\infty$, 
$N_{\rm sk,\Omega}=-1$ for a skyrmion on $\Omega$.  

\begin{figure}[t]
\centering\includegraphics[width=\linewidth,angle=0,clip]{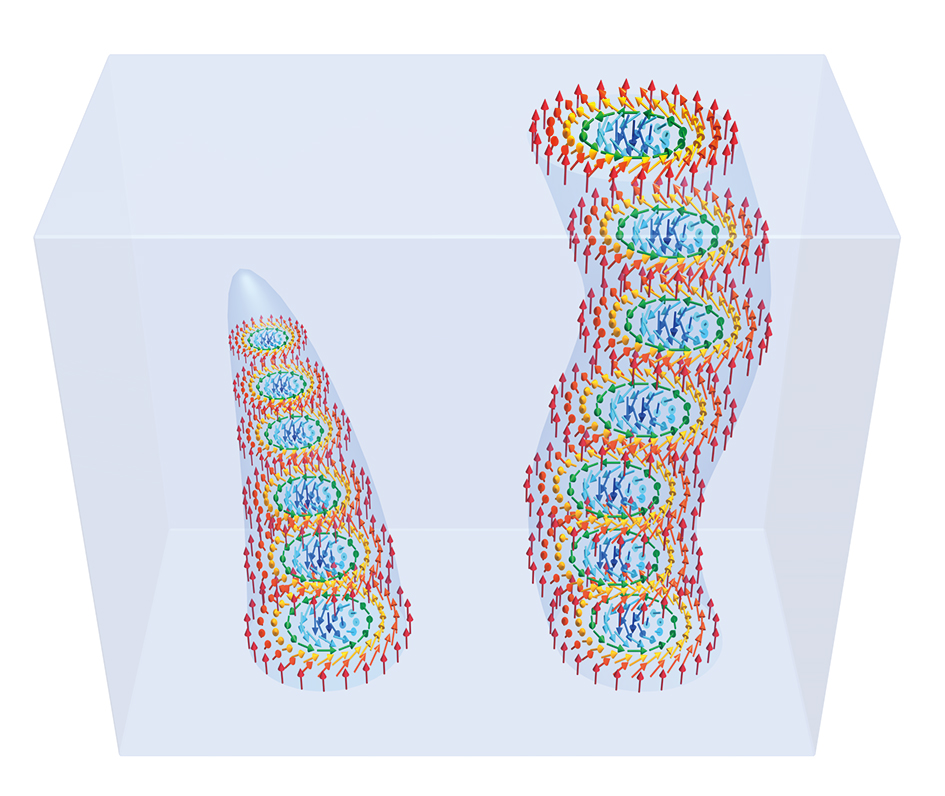}
\caption{\label{fig1}
Skyrmion string ending with the skyrmion on the surface (right) and that terminates at the monopole (left).
}
\end{figure}

In the three dimensional magnets, the skyrmion usually forms rod-like object along 
the external magnetic field~\cite{Schuette,Milde,Lin,Rybakov,Kanazawa1,Kanazawa2,Kagawa,ShileiZhang,Birch}. When the meandering degree of freedom is introduced, it is better to consider it as skyrmion string. 
When the skyrmion string terminates or branches into two skyrmion strings in the bulk, 
the singular points appear.  
(Figure \ref{fig1} is a schematic for the skyrmion string (right) and that with a singular point (left).) 
The study of such singular points goes back over more than 
a half-century~\cite{Feldtkeller,Slonczewski,MalozemoffSlonczewski,Chikazumi,Braun}.  
In those earlier studies~\cite{Feldtkeller,Slonczewski,MalozemoffSlonczewski,Chikazumi}, the Bloch point, namely,  the topological defect on the Bloch line was extensively studied.  
The topologically the same defects are sometimes called 
(anti)hedgehog or (anti)monopole~\cite{Schuette,Milde,Kanazawa1,Kanazawa2,Braun}.  
In the present paper, we use the word, (anti)monopole, 
to express the topological defect on the skyrmion string. 

%%%%%%%%%%%%%%%%%%%%%%%%%%%%%%%%%%%%%%%%%%%%%%%%%
%%%%%%%%%%%%%%%%%%%%%%%%%%%%%%%%%%%%%%%%%%%%%%%%

Kotiuga~\cite{Kotiuga1,Kotiuga2} described the topological nature of (anti)monopole 
by the Hopf extension theorem of algebraic topology.  
It is nothing but the Gauss' low for the topological charge and flux:
The (anti)monopole is characterized by the topological index $N_{\rm mp}$ called monopole charge. This $N_{\rm mp}$ is defined by the integral of the solid angle formed by the magnetic moments over the surface enclosing the (anti)monopole:  
Using Eq.~(\ref{eqn:topo}), the monopole charge is defined as 
$N_{\rm mp}=N_{\rm topol}(\Omega_{\bm r_\text{mp}})=+1$ 
($N_{\rm mp}=N_{\rm topol}(\Omega_{\bm r_\text{amp}})=-1$)
for 
$\Omega_{\bm r_\text{mp}}$ ($\Omega_{\bm r_\text{amp}}$ )
enclosing a monopole at $\bm r_\text{mp}$ 
(an antimonopole at $\bm r_\text{amp}$) with ${\bm e}$ pointing outward 
the domain of integral.  
For a closed surface $\Omega$ which does not 
enclose the spatial defects such as void(s), Eq.~(\ref{eqn:topo}) gives 
\begin{align}
N_{\rm sk,\Omega}
&=\sum_{\bm r_{\rm mp}}N_{\rm mp}(\bm r_{\rm mp})
  +\sum_{\bm r_{\rm amp}}N_{\rm mp}(\bm r_{\rm amp})\notag \\
&=\sum_{\bm r_{\rm mp}}+1\ \ \ +\ \ \ \sum_{\bm r_{\rm amp}}-1,
\label{eqn:Gauss} 
\end{align}
for (anti)monopoles enclosed in $\Omega$ and (anti)skyrmions on $\Omega$. 
For the flux density $\frac{1}{2\pi}\bm b$, this Gauss' law relates the skyrmion string and the (anti)monopole, i.e., those are corresponding to the flux line and its source (sink) point.  
The total monopole charge for the (anti)monopoles enclosed by $\Omega$ 
is always the same as the total skyrmion number 
on the surface $\Omega$, $N_{\text{sk},\Omega}$.

In some cases, the endpoints of a skyrmion string on the surface of magnet 
might be regarded as the monopole and antimonopole. 
However, $N_{\rm mp}$ cannot be defined for the surface magnetic texture 
since half of the space is ``vacuum'' where magnetic moment is absent. 
In particular, the (anti)monopole point $\bm r_\text{mp}$ ($\bm r_\text{amp}$)
defined above cannot be on the surface of magnet.  
On the other hand, one can define $N_{\rm sk,\Omega}=N_{\rm topol}(\Omega)$ for the magnetic texture on the surface $\Omega$, i.e., in this case, 
the surface of a magnet gives a well-defined orientable two-dimensional manifold $\Omega$.  

The topological nature discussed above is essential to discuss the stability of 
the magnetic texture.  For the magnetic moments on a two-dimensional lattice, the topological stability is based on the energy scales of the excitation. For example, in the chiral magnets with the ferromagnetic interaction $J$ and Dzyaloshinskii-Moriya (DM) interaction $D$~\cite{Dzyaloshinskii,Moriya1,Moriya2}, 
the length scale of the skyrmion size is characterized by 
$\sim$$(J/D)a$ with the lattice constant $a$, which is much larger than $a$ when $D\ll J$. This fact validates the continuum approximation, and the energy density is 
$\sim$$(D^2)/(Ja^2)$.  
This energy density and the skyrmion size result in the order of $J$ for 
the energy scale of the stability for a skyrmion.  
Therefore, a change in $N_{\rm sk}$, i.e., the topological transition of magnetic texture requires an overcome of the energy barrier of the order of $J$.  
When a skyrmion string is broken at a point $\bm r_b$ in bulk, 
a monopole-antimonopole pair appears at the point.   
In other words, at the two-dimensional cross section $\Omega$ 
including the broken point $\bm r_b$, 
the skyrmion number $N_{\rm sk,\Omega}$ changes. 
Therefore, this change also requires 
the overcome of the energy barrier of the order of $J$.  

In the present paper, we show that the surface $N_{\rm sk}$ plays a crucial role together with $N_{\rm mp}$ for the skyrmion string stability and dynamics.  
To this end, we numerically investigate the current driven dynamics of the skyrmion string in the magnet with step edges on the surface. 
The step edges act as the pinning center of the motion of a skyrmion string, which sometimes leads to the detachment of the skyrmion from the surface or the splitting of the string into pieces. 
By the numerical simulation, we examine the stability of the surface (anti)skyrmion and the dynamics including (anti)skyrmion-(anti)monopole collision leading to skyrmion string annihilation. 
These stability and dynamical processes are well understood as two kinds of topological indices; skyrmion number $N_{\rm sk}$ for the surface and the monopole charge $N_{\rm mp}$ for the bulk.

\vskip0.5cm
\noindent
{\bf Results}

\begin{figure*}[t]
\centering\includegraphics[width=14cm,angle=0,clip]{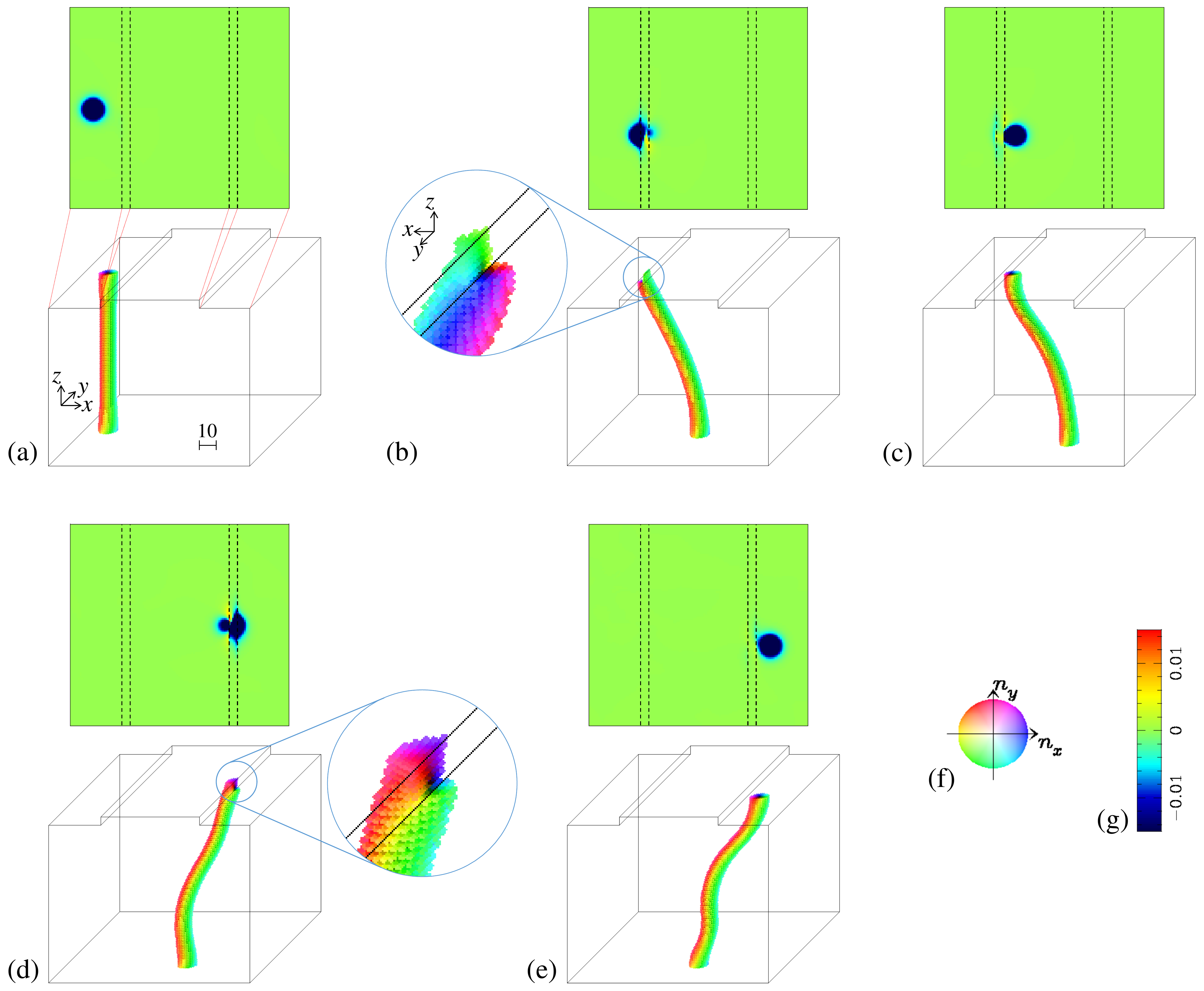}
\caption{\label{fig2}
Skyrmion string in chiral magnets with step edges 
under an external magnetic field along $z$-direction. 
The case with step height 5 is shown. 
(a) The upper (lower) panel represents the spatial distribution of emergent $b$-field $b_\text{normal}$ normal to the surface 
using color code (g) (magnetic texture using color code (f)) at $t=0$.  
The color code (f) indicates $n_x$-$n_y$ component of the magnetic moments, e.g., 
blue is corresponding to the in-plane magnetic moment along $x$ axis.   
The darkness of the color represents the $n_z$ component, i.e., black is corresponding to $\bm n=(0,0,-1)$.  
The broken lines in the upper panel are corresponding to 
the edges of the upper and lower terraces as indicated by the red dotted lines.  
In the same way, the snapshots at (b) $t=6000$, 
(c) $t=8040$, (d) $t=11060$ and (e) $t=11500$ are shown.  
In (b) and (d), the enlarged magnetic textures at the step edges are shown.  
(For (b), the arrangement of the magnetic texture is seen from the left side. ) 
To make the visualization of string clear, the magnetic moments with $n_z>0.5$ are not shown for the panels of magnetic texture.   
}
\end{figure*}

To study the topological stability of (anti)skyrmion, (anti)monopole and skyrmion string, we start with a metastable skyrmion string in 
a three-dimensional chiral magnet with step edges 
(see Fig.~\ref{fig2}).  
The Hamiltonian is given by  
\begin{align}
\mathcal{H}=\sum_{\bm r\in\Lambda}E(\bm r), 
\label{eqn:Hamiltonian}
\end{align}
with
\begin{align}
E(\bm r)=&\sum_{\bm r+{\bm \rho} \in \Lambda}
\frac{1}{2}\left[-J{\bm n}_{\bm r} \cdot {\bm n}_{{\bm r}+{\bm \rho}} 
+D\left({\bm n}_{\bm r} \times {\bm n}_{{\bm r}+{\bm \rho}} 
\right)\cdot {{\bm \rho}}
\right]
\notag \\
&-h  n_{z,\bm r},
\label{eqn:Hamlocal}
\end{align}
where ${\bm \rho}=\pm\hat{\bm x}, \pm\hat{\bm y}, \pm\hat{\bm z}$ with   
the unit vectors $\hat{\bm x}$, $\hat{\bm y}$ and $\hat{\bm z}$ 
in $x$-, $y$- and $z$- axes, 
and $\Lambda$ is the set for the cubic lattice sites of the system.  
The normalized magnetic moments at $\bm r\in \Lambda$ is denoted by 
$\bm n_{\bm r}=(n_{x,\bm r},n_{y,\bm r},n_{z,\bm r})$.  
The lattice constant is taken as the unit of length.  
As shown in Fig.~\ref{fig2}, the step edges are introduced on the top surface of the magnet while the bottom surface is flat. 
The step edge is perpendicular to $x$-direction.  
In $x$- and $y$- directions, the periodic boundary condition is imposed.  
But, where the bottom surface and the top surface with step edges face to ``vacuum'', the open boundary condition is employed.   
For simulations, we use the system size with 
$120\times120$ for the bottom surface area 
and $z=1\sim100$ at the higher terrace area.  
The higher terrace has a width 60.  
Figure \ref{fig2} shows a case with a step height 5, 
i.e., the lower terrace is on the layer with $z=95$.  

Here, we use a parameter set $\{J=1, D=0.2, h=0.06\}$ (i.e., $J$ is the unit of $D$ and $h$) where the ferromagnetic state polarized in $z$ direction 
is the ground state~\cite{Han1,Mochizuki}. 
Figure \ref{fig2}(a) is the relaxed metastable state with a skyrmion string which is in the 
lower terrace area.  
The skyrmion string has a tensile strain due to the metastablity, 
i.e., the longer string costs more energy.  
Therefore, the relaxed string is straight along $z$ direction.  
Consequently, the string in the higher terrace area has an energy cost 
compared to the string in the lower area. 
In other words, the height profile of this system roughly indicates the potential profile for the skyrmion string. (See also Supplementary Information.)

We drive the skyrmion string by the 
spin-transfer-torque (STT) effect~\cite{NagTok}: 

\noindent 
The Landau-Lifshitz-Gilbert (LLG) equation is given by
\begin{align}
\frac{{\rm d} {\bm n}_{\bm r}}{{\rm d} t}=& 
- \frac{\partial \mathcal{H}}{\partial {\bm n}_{\bm r}}
\times {\bm n}_{\bm r} 
+\alpha {\bm n}_{\bm r} \times 
\frac{{\rm d}{\bm n}_{\bm r}}{{\rm d} t}\notag\\
&-\left(\bm j\cdot\nabla\right)\bm n_{\bm r}
+\beta\left[\bm n_{\bm r}\times\left(\bm j\cdot\nabla\right)\bm n_{\bm r}\right],
\label{eqn:LLG}
\end{align}
where $\alpha$ is the Gilbert damping constant.  
The last two terms in Eq.~(\ref{eqn:LLG}) represent 
the STT effect due to the spin polarlized 
electric current density $\bm j$ with 
the coefficient of the non-adiabatic effect $\beta$. 
In the following, we examine the skyrmion dynamics for 
the current $j=|\bm j|=0.006$ parallel to $\hat{\bm x}$ under 
the condition $\alpha=\beta$ (=0.01) to avoid its current driven Hall motion.

%%%%%%%%%%%%%%%%%%%%%%%%%%%%%%%%%%%%%%%%%%%%%%%%%
%%%%%%%%%%%%%%%%%%%%%%%%%%%%%%%%%%%%%%%%%%%%%%%%

%%%%%%%%%%%%%%%%%%%%%%%%%%%%%%%%%%%%%%%%%%%%%%%%%
\begin{figure*}[t]
\centering\includegraphics[width=15cm,angle=0,clip]{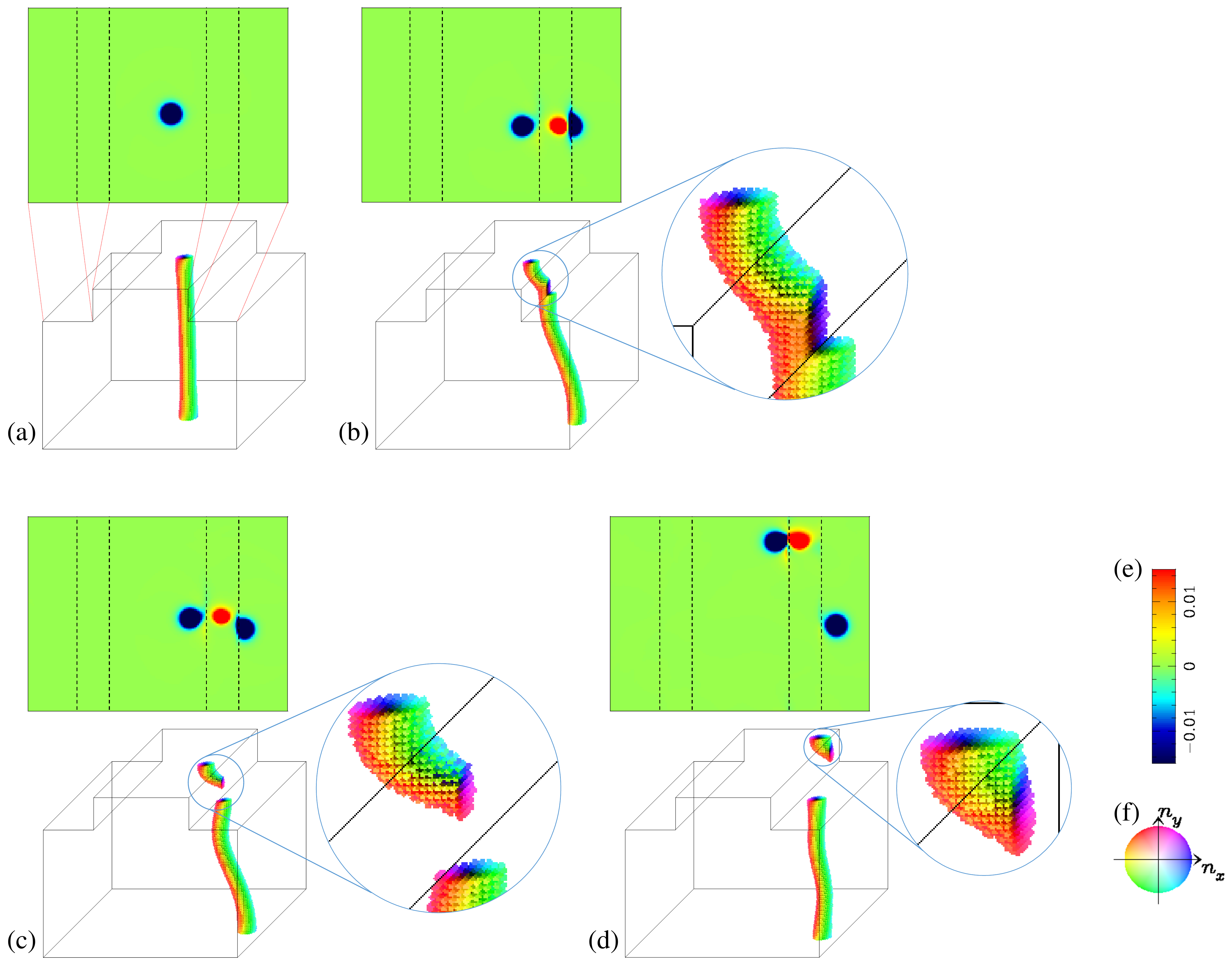}
\caption{\label{fig3}
Pair creation and annihilation of the surface (anti)skyrmion(s).  
The height of the step edge is 20.  
(a) The upper (lower) panel represents the spatial distribution of emergent $b$-field $b_\text{normal}$ normal to the surface using color code (e) (magnetic texture using color code (f)) at $t=0$. 
The broken lines in the upper panel are corresponding to 
the edges of the upper and lower terraces as indicated by the red dotted lines.  
In the same way, the snapshots at (b) $t=4000$, 
(c) $t=4800$, and (d) $t=5400$ are shown.  
In (b), (c) and (d), the enlarged images of the magnetic textures at around the right edge 
are also shown. }
\end{figure*}
%%%%%%%%%%%%%%%%%%%%%%%%%%%%%%%%%%%%%%%%%%%%%%%%

\vskip0.5cm
\noindent
{\bf Skyrmion number for surface magnetic texture}

In this section, we first  
show the stability of the (anti)skyrmion at the surface, i.e., 
the surface (anti)skyrmion is not easy to annihilate even in the presence of the step edges 
of moderate height (the case summarized in Fig.~\ref{fig2}). 
It is also shown by the conservation of the skyrmion number at the surface.  
Next, it is shown that 
the conservation of the skyrmion number at the surface applies for more complex dynamics 
where the skyrmion string is separated into pieces due to the large height of step edge (the case summarized in Fig.~\ref{fig3}).

Figure \ref{fig2} summarizes 
the current driven dynamics of the skyrmion string 
in the system with the step height 5.  
By the STT effect, the skyrmion string approaches to the left step edge 
(see Figs.~\ref{fig2}(a)$\to$(b)).  
However, the edge prevents the motion of the top endpoint of the string.  
Deep inside the magnet, the string moves by the STT effect, and gets bent and stretched as seen in Fig.~\ref{fig2}(b).   
After that, the top endpoint of the string, i.e., the top surface skyrmion  
overcomes the pinning due to the step edge and 
climbs up to the higher terrace as seen in Figs.~\ref{fig2}(a)$\to$(b)$\to$(c).  
After that, the skyrmion string shows a characteristic dynamics~\cite{KoshibaeSR2} 
like `moving tornado' reflecting the vorticity, Magnus effect and the tensile strain.  
(See also Supplementary Information and movie S3.avi.)    
The current driven skyrmion string approaches to the right step edge and 
the top surface skyrmion at the higher terrace goes down to the lower terrace 
as seen in Figs.~\ref{fig2}(c)$\to$(d)$\to$(e).  
An interesting aspect of the dynamics is that the upper endpoint of the string always 
sticks to the surface even when the surface bents with 90 degree at the step edge. 
Figures \ref{fig2}(b) and (d) actually shows the behaviors of the top surface skyrmion. 
(See also Supplementary Information).   

This is understood to be the topological stability of the skyrmion at the surface:  
A way to define the topological nature of the string might be 
\begin{align}
N_{\rm sum}=\sum_z N_{\rm topol}(\Omega_z)
\label{eqn:sksum}
\end{align} 
with $\Omega_z$ being the horizontal plane at height $z$ and 
${\bm e}=\hat{\bm z}$.  
In the present case, there exist two regions with different heights of the top surface due to the step edges.  
Accordingly, $N_{\rm sum}$ changes along 
the dynamics, e.g., Figs.~\ref{fig2}(a)$\to$(b)$\to$(c) and Figs.~\ref{fig2}(c)$\to$(d)$\to$(e).  
However, the change in magnetic texture along 
the dynamics shown in Fig.~\ref{fig2} occurs within a continuous deformation 
without topological singularity.  Therefore, $N_{\rm sum}$ cannot be appropriate for the topological index for the magnetic texture.    
On the other hand, when we define the skyrmion number 
\begin{align}
N_{\rm sk, top}=N_{\rm topol}(\Omega_\text{top})
\label{eqn:sktop} 
\end{align}
with $\Omega_\text{top}$ being the developed top surface and 
${\bm e}$ points outward the magnet,     
it is confirmed that $N_{\rm sk, top}$ is conserved during the dynamics 
summarized in Fig.~\ref{fig2}.  At the same time,     
it represents the topological protection of the magnetic texture 
at the top endpoint of the string.   
The topologically protected surface skyrmion dynamics is also  
well described by the time evolution of the spatial distribution of 
the emergent $b$-field $b_\text{normal}$ normal to the surface $\Omega_\text{top}$ 
which directly probes the deformation of the skyrmion (see the top panels of 
Figs.~\ref{fig2}(a)-(e). 
Although the skyrmion is strongly deformed due to the steep structure at step edges, 
the skyrmion keeps stick to the top surface during the dynamics. 
(See also Supplementary Information.)

We find that the surface topological index $N_{\rm sk, top}$ 
is applicable for more complex phenomenon.  
Figure \ref{fig3} shows the skyrmion string dynamics in the system with 
the step edges of height 20 and the string starts in the higher terrace area.  Other conditions are the same as those for the case Fig.~\ref{fig2}.  
By the STT effect, the string approaches the right step edge.  Because of the repulsive interaction between the right step edge and the string, the string shows a bending 
behavior and touches the lower step corner first whereas its upper endpoint is still away from the step edge (see Fig.~\ref{fig3}(b)).  At the same time, at around the touched point, the magnetic texture of the string shows a deformation and finally the string splits into two parts, as shown in Figs.~\ref{fig3}(b)$\to$(c).  
Note that the endpoint at the right step edge of the shorter string 
has a positive contribution to the topological index $N_{\rm sk, top}$ 
whereas the contribution by the endpoint at the higher terrace surface is negative 
(see the plot of $b_\text{normal}$ in Figs.~\ref{fig3}(b)$\to$(c)$\to$(d).  
In other words, the emergent magnetic texture at the right step edge is the antiskyrmion. 
This causes a characteristic dynamics due to the topological nature~\cite{KoshibaeNJP,KoshibaeSR}: 
After the skyrmion-antiskyrmion pair-creation shown as Figs.~\ref{fig3}(a)$\to$(b)$\to$(c), 
the skyrmion-antiskyrmion pair, i.e, the endpoints of the shorter string run together in $+\hat{\bm y}$ direction as seen in Figs.~\ref{fig3}(c)$\to$(d).   
The (anti)skyrmion has a vorticity and its sign is consistent with 
the sign of the topological index. 
Because of the vorticity, a Magnus force appears perpendicular to the force acting on the 
(anti)skyrmion~\cite{KoshibaeNJP,KoshibaeSR}.  
In the present case, due to the tensile strain on the shorter string, 
an attractive force is acting on the endpoints of the string, i.e., 
the skyrmion at the higher terrace and the antiskyrmion at the right step edge of the 
top surface.  
Since the vorticity of the skyrmion is opposite to that of the antiskyrmion, 
the attractive force drives the Magnus force for the skyrmion and the antiskyrmion 
in the same direction. 
With this dynamics, finally, the shorter string disappears with 
the skyrmion-antiskyrmion pair annihilation at the top surface.  
Note that during the time evolution summarized in Fig.~\ref{fig3}, 
$N_{\rm sk, top}$ is conserved. 
This dynamics occurs without singularity of the magnetic configuration.   

%%%%%%%%%%%%%%%%%%%%%%%%%%%%%%%%%%%%%%%%%%%%%%%%%
\begin{figure*}[t]
\centering\includegraphics[width=13cm,angle=0,clip]{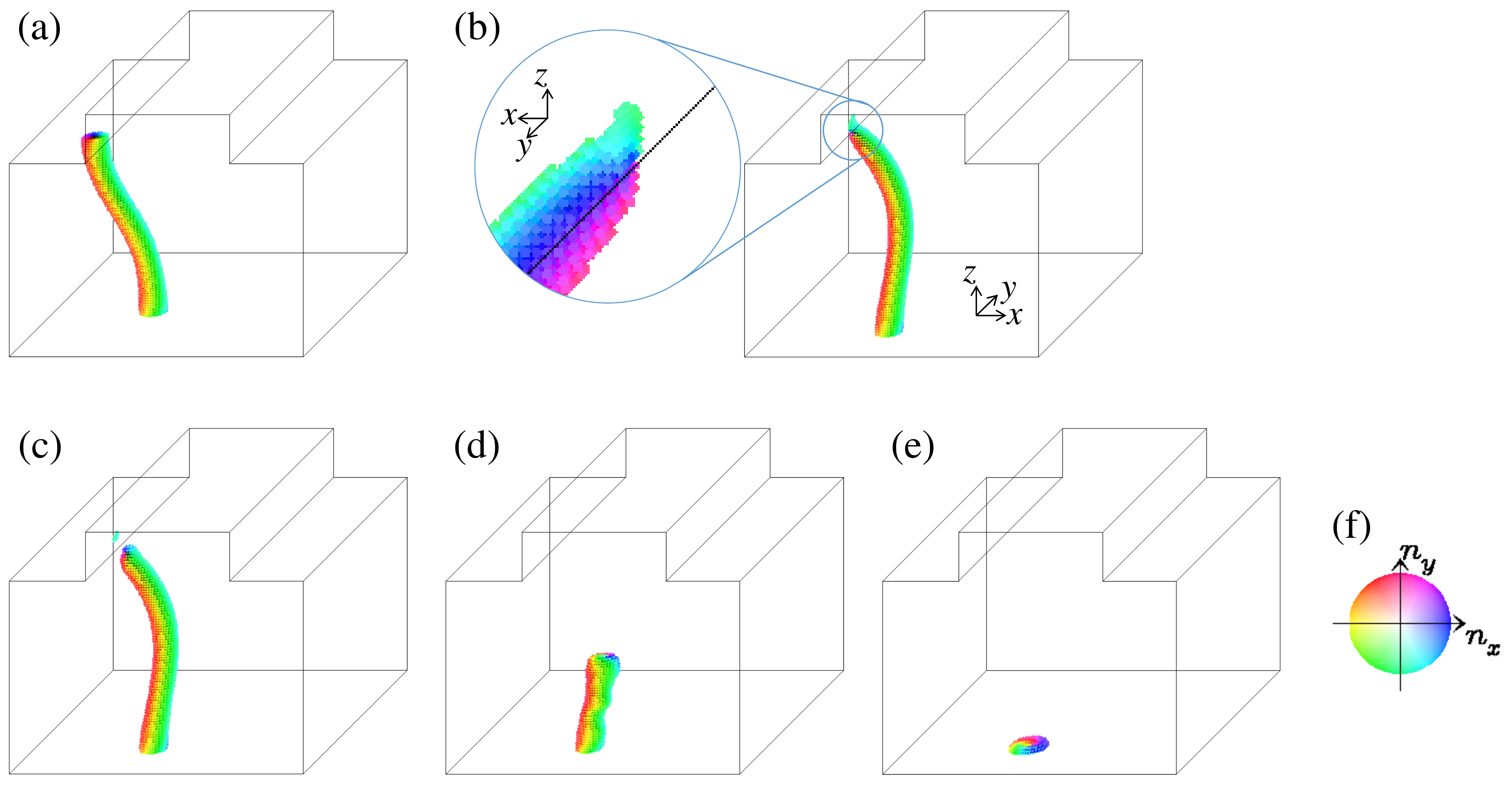}
\caption{\label{fig4}
Monopole dynamics.  The height of the step edge is 20.  
Snapshots of the magnetic texture at (a) $t=14000$, (b) $t=16290$, (c) $t=16300$, (d) $t=16400$ and (e) $t=16480$ are shown. (See text.) 
In (b), the enlarged magnetic texture is seen from left.  
(f) The color code for the magnetic texture.}
\end{figure*}
%%%%%%%%%%%%%%%%%%%%%%%%%%%%%%%%%%%%%%%%%%%%%%%%

%%%%%%%%%%%%%%%%%%%%%%%%%%%%%%%%%%%%%%%%%%%%%%%%%
\begin{figure}[b]
\centering\includegraphics[width=\linewidth,angle=0,clip]{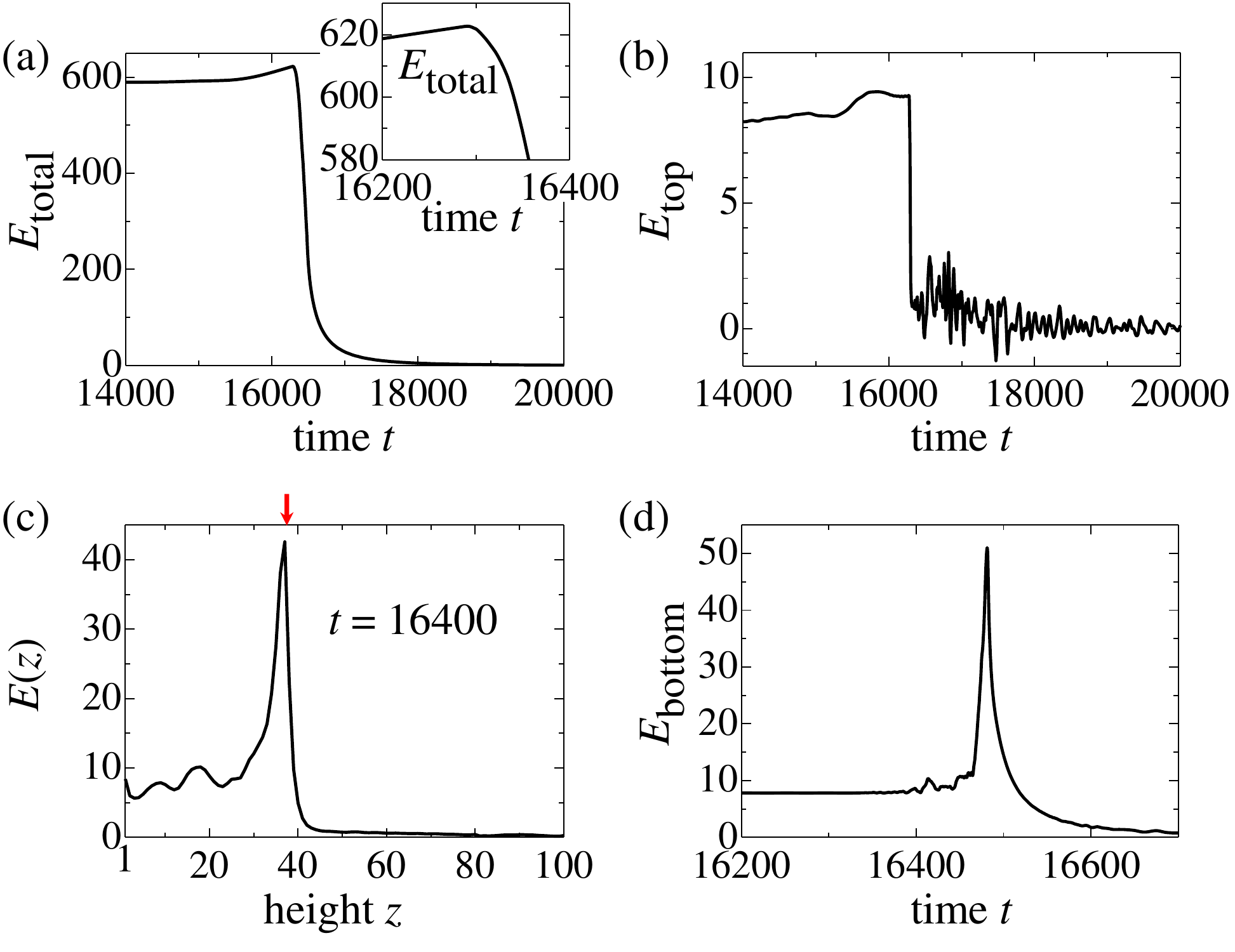}
\caption{\label{fig5}
Monopole and singularity. 
(a) Time dependence of the total energy.  The enlarged plot at around $t\sim16300$ is also presented.  
(b) Time dependence of the energy of the top surface.  
(c) The local energy at height $z$.  The red arrow indicates $z_\text{mp}$.  The sharp decrease occurs at the time when the skyrmion is detached. (See text).
(d) Time dependence of the energy of the bottom surface.  
The peak occurs at the time where the skyrmion on the bottom surface disappears.}
\end{figure}
%%%%%%%%%%%%%%%%%%%%%%%%%%%%%%%%%%%%%%%%%%%%%%%%

\vskip0.5cm
\noindent
{\bf Creation and annihilation of monopole}

In the present system, we can also discuss the magnetic texture with singularity of the  magnetic configuration. 
Figure \ref{fig4} summarizes the skyrmion string dynamics 
after those shown in Fig.~\ref{fig3}.  
The STT effect drives the string in $\hat{\bm x}$ direction.  
(Note that we impose the periodic boundary condition along $x$- and $y$-directions.)
Similar to the dynamics in the initial stage shown in Fig.~\ref{fig2}, 
the skyrmion at the upper endpoint of the string sticks to the top surface.  
(See Fig.~\ref{fig4}(a) and (b).)  
However, because the height of the step edge is high enough, the skyrmion string cannot overcome the barrier, and the skyrmion on the top surface is detached 
(Figure \ref{fig4}(c) is the magnetic texture just after this `detach' event.)   
After that, this upper endpoint of the string runs along the string and finally 
the string totally disappears. 
(See Supplementary Information.)  
After the upper endpoint of the string is detached, 
we find the topological discontinuity, 
i.e., the emergence of monopole:
The monopole point $\bm r_{\rm mp}=(x_{\rm mp},y_{\rm mp},z_{\rm mp})$ 
is an accumulation point where the magnetic moment is ill-defined. Therefore, it is not on the lattice site in $\Lambda$.  
Using $N_\text{topol}(\Omega_z)$, the topological discontinuity by $\bm r_{\rm mp}$ is expressed to be, 
$N_\text{topol}(\Omega_z)=-1$ for $z<z_{\rm mp}$ and 
$N_\text{topol}(\Omega_z)= 0$ for $z>z_{\rm mp}$ and a relevant definition of $(x_{\rm mp},y_{\rm mp})$ will be given by the minimum of $n_z$ with an interpolated function on the the horizontal plane at $z=z_{\rm mp}$.

\vskip0.5cm
\noindent
{\bf Topological indices} 

The monopole charge is defined by 
$N_{\rm mp}=N_\text{topol}(\Omega)$  
with $\Omega$ enclosing $\bm r_{\rm mp}$ as discussed 
in the paragraph with Eq.~(\ref{eqn:Gauss}).  
For the numerical results summarized in Figs.~\ref{fig2}$\sim$\ref{fig4}, we find that the following relation always holds,  
\begin{align}
N_{\rm mp}=N_{\rm sk, top}+N_{\rm sk, bottom},
\label{eqn:tousiki} 
\end{align}
where $N_{\rm sk, bottom}=N_\text{topol}(\Omega_\text{bottom})$ and 
$\Omega_\text{bottom}$ is the bottom surface with 
${\bm e}=-\hat{\bm z}$.  (See also Eq.~(\ref{eqn:sktop}).)  
Note that $N_{\rm sk, bottom}=-N_\text{topol}(\Omega_z)$ with 
$z=1$ (see Eq.~(\ref{eqn:sksum}), and here, $z=1$ represents the bottom of the magnet). 
The domain of integral $\Omega$ enclosing $\bm r_{\rm mp}$ is topologically the 
same as $\Omega_\text{top}+\Omega_\text{bottom}$. 
During the process shown in Fig.~\ref{fig2} and Fig.~\ref{fig3}, 
$N_{\rm sk, top}$ and $N_{\rm mp}$ are always zero.  
At the detach process of the top surface endpoint 
of the skyrmion string described in Fig.~\ref{fig4}, 
the simultaneous change $N_{\rm sk, top}=-1\to0$ and $N_{\rm mp}=0\to+1$ occurs 
(and $N_{\rm sk, bottom}=+1$ is kept).

For the dynamics summarized in Fig.~\ref{fig4}, let us discuss the relation between the topological characteristics discussed above and  the metastabilities of the magnetic textures, in more detail.  
Figure \ref{fig5}(a) shows the time dependence of the total energy $E_\text{total}$ 
measured from that of the relaxed ferromagnetic state.  
(See Supplementary Information.)   
Along the dynamics Fig.~\ref{fig4}(a)$\to$(b), the total energy $E_\text{total}$ increases.  
This is because the upper endpoint is pinned by the left step edge and the 
string is bent and stretched by the STT effect. After that the total energy $E_\text{total}$ decreases rapidly  
with the detach of the top surface endpoint and 
successively occurring monopole dynamics Fig.~\ref{fig4}(b)$\to$(c)$\to$(d). 
At the detach process, the total energy $E_\text{total}$ shows rather smooth time dependence.  
The `hidden' singular behavior along the emergence of the monopole is observed 
in the time dependence of the local energy at the top surface: 
We define the local energy on $\Omega$ by   
\begin{align}
E(\Omega,t)=\sum_{\bm r\in\Omega}\left[E_t(\bm r)-E_f(\bm r)\right], 
\label{eqn:locale}
\end{align}
where $E_t(\bm r)$ ($E_f(\bm r)$) is 
given by Eq.~(\ref{eqn:Hamlocal}) for the instantaneous magnetic texture at time $t$ 
(for the relaxed ferromagnetic texture).

The plot Fig.~\ref{fig5}(b) shows $E_\text{top}=E(\Omega_\text{top},t)$ 
as a function of time $t$.  
At around $t\sim14000$, $E_\text{top}$ hardly changes because the skyrmion at the top surface is apart from the step edge.  With approaching the skyrmion to the (left) step edge 
by the STT effect, the skyrmion becomes unstable due to its deformation. This causes the increase of $E_\text{top}$ and finally 
the sharp drop of $E_\text{top}$ occurs at the time when the skyrmion is detached, 
i.e., the emergence of the monopole.  The increase of $E_\text{top}$ before the emergence of the monopole indicates the energy barrier dividing the skyrmionic state and ferromagnetic state at the top surface $\Omega_\text{top}$.  The profile of the energy barrier seen in Fig.~\ref{fig5}(b) is rather moderate compared to that in the discussion below because of the geometry with the step edge, i.e., the steep geometrical arrangements of the top surface reduce the metastability of the top surface skyrmion.

The singularity of the monopole is obvious in the local energy profile as shown  in Fig.~\ref{fig5}(c).  
This plot shows the height $z$ (=1$\sim$100) dependence of 
$E(z)=E(\Omega_z,t=16400)$.  
We clearly see the sharp energy peak which divide the metastable skyrmionic state and the ferromagnetic state.  The red arrow on the top horizontal axis indicates the position 
$z_\text{mp}$, i.e., it divides the region of $z$ by $N_\text{topol}(\Omega_z)=-1$ or 0.  
(See also Supplementary Information and movie S8.avi.)

In Fig.~\ref{fig5}(a), after the monopole creation, 
the total energy $E_\text{total}$ decrease as a function of time $t$ 
smoothly, i.e., no singular behavior is seen.  
This indicates a smooth motion of the monopole which
makes the metastable skyrmion string shorter, 
although the monopole is a singular object as seen in Fig.~\ref{fig5}(c).

At the final stage, the collision of the monopole and the antiskyrmion occurs, and 
the monopole, the antiskyrmion on the bottom and the skyrmion string 
totally disappear with the simultaneous change of $N_{\rm sk, bottom}$ and $N_{\rm mp}$ from +1 to 0.  
When we focus only on the bottom surface, we see the singularity with the energy cost: 
Figure \ref{fig5}(d) shows the time dependence of the local energy at the bottom surface, 
$E_\text{bottom}=E(\Omega_\text{bottom},t)$.  
The sharp peak structure occurs 
with the simultaneous change of $N_{\rm sk, bottom}$ and $N_{\rm mp}$ from +1 to 0.  
However, in the total energy $E_\text{total}$, this energy cost is compensated by the annihilation of the skyrmion string in total.

\vskip0.5cm
\noindent
{\bf Discussion and summary}

The Gauss' law Eq.~(\ref{eqn:Gauss}) 
applies for the processes discussed in the present paper:  
Suppose $\Omega$ is the whole surface of the magnet and 
the magnet has no spatial defects such as voids. 
There are two cases,  (A) div~$\bm b=0$ in bulk and (B) div~$\bm b\neq 0$ in bulk.

\noindent
- In case (A), the system has no (anti)monopoles.  
As shown in Fig.~\ref{fig3}, the skyrmion string is divided into two within the 
continuous deformation of the magnetic texture.  
As a result, using the surface $\Omega$, 
any entanglements of the skyrmion string even in the presence of the knots, 
are solved without topological transitions. 
Therefore, it is concluded that any skyrmionic states 
are homeomorphic to each other and also those are topologically the
same as ferromagnetic states and helix states in bounded 
three-dimensional magnets.

\noindent
- In case (B), the system has (anti)monopoles.  
The (anti)monopole is a topologically singular object and cannot be created/annihilated 
within the continuous deformation of the magnetic texture.  
For a monopole-antimonopole pair,  
$N_{\rm mp}(\bm r_{\rm mp})+N_{\rm mp}(\bm r_{\rm amp})=0$ and it does 
not contribute to Eq.~(\ref{eqn:Gauss}). 
Therefore, Eq.~(\ref{eqn:Gauss}) is not appropriate to describe the topological invariance for the magnetic texture on the whole system.

To discuss the stability of the magnetic textures, the ``local'' monopole charge is important. 
The (anti)monopole always accompanies the high energy (being order of $J$) 
area concentrated at around $\bm r_{\rm mp}$ ($\bm r_{\rm amp}$).  
Consequently, for example, to break a skyrmion string into two 
at the point deep inside the magnet, 
for the monopole-antimonopole pair creation in other words, 
a large energy to overcome the energy barrier being order of $J$ is required~\cite{KoshibaeSR2}.  
In this case, the change in absolute value 
$|N_{\rm mp}(\bm r_{\rm mp})|+|N_{\rm mp}(\bm r_{\rm amp})|$ is important 
rather than total monopole charge.

The energy cost at the (anti)monopole creation/annihilation is compensated by  
the shrinkage/deformation of the skyrmion string connecting the (anti)monopole 
as seen in Figs.~\ref{fig4} and \ref{fig5}.    
At the detach process of the skyrmion string from the top surface 
shown in Figs.~\ref{fig4}(b)$\to$(c), 
we calculate the skyrmion number 
$N_\text{sk,next-to-top}=N_{\rm topo}(\Omega_\text{next-to-top})$ 
where $\Omega_\text{next-to-top}$ is the top surface of 
$\Lambda-\Omega_\text{top}$ ($\Lambda$ is the set of all sites $\bm r$ 
of the system defined below Eq.~(\ref{eqn:Hamlocal})).  
We find a time duration with $N_\text{sk,top}=0$ and 
$|N_\text{sk,next-to-top}|=1$.  
This means that the monopole point $\bm r_\text{mp}$ appears as an 
accumulation point between $\Omega_\text{top}$ and $\Omega_\text{next-to-top}$.  
Therefore, the monopole point $\bm r_\text{mp}$ emerges without change of the length of the skyrmion string essentially, so that  
the energy cost due to the energy barrier discussed above appears 
in the time dependence of the total energy as seen in Fig.~\ref{fig5}(a).  
Even so, the sharp singularity due to the emergence of the monopole $point$ 
is smeared in the total energy in three dimension.

In the present paper, we have seen the importance of the topological indices 
$N_\text{sk}$ and $N_\text{mp}$.  These indices, specifically, 
are related by the Gauss' law Eq.~(\ref{eqn:Gauss}).  
In the previous studies~\cite{Kagawa,KoshibaeSR2}, 
it is discussed that 
the monopole dynamics running through the string 
causes the skyrmion string annihilation. 
The annihilation of a skyrmion string is seen 
in the final stage of the dynamics in Fig.~\ref{fig4}, i.e., 
the collision of the monopole and the antiskyrmion at the bottom surface.  
On the dynamics, the skyrmion number at the bottom surface 
$N_\text{sk,bottom}$ changes from +1 to 0. 
As seen in Fig.~\ref{fig5}(d), a steep enhancement of $E_\text{bottom}$ occurs 
with the change of $N_\text{sk,bottom}$.  
However, this enhancement of $E_\text{bottom}$  
does not result in the protection of the bottom surface antiskyrmion.  
The energy cost by the 
local topological singularity seen in Fig.~\ref{fig5}(d) is totally compensated by 
the energy gain due to the shrinking of the skyrmion string.  
Consequently, on the time window of this monopole-antiskyrmion collision dynamics, 
the total energy (see Fig.~\ref{fig5}(a)) decreases smoothly and monotonously.  
Note that the change of the topological indices 
$N_\text{sk}$ and $N_\text{mp}$ occurs at the same time, and 
the Gauss' law Eq.~(\ref{eqn:Gauss}) always holds along the dynamics discussed here.

The skyrmion string annihilation instability is responsible for the (anti)monopole dynamics. 
For shorter skyrmion string, 
the probability of the emergence of the (anti)monopole(s) is reduced.  
This is why the skyrmion string is more stable in thinner magnets~\cite{YuXZNM11}.

To summarize, we have discussed topological particles and strings on the magnets and their characteristic dynamics, e.g., particle-antiparticle pair creation/annihilation, collisions of the particles and behind string dynamics.  
To describe the dynamical processes of skyrmion string, (ant)skyrmion and (anti)monopole,  we have shown that two topological indices, i.e., $N_{\rm sk}$ on the surface and $N_{\rm mp}$ in the bulk play the essential role.

\vskip0.5cm
\noindent
{\bf Methods}

The units of time $t$ is $1/J$.  
Typically $J \sim 10^{-3}$ eV and the unit $1/J$ becomes 
$\sim$0.7 ps.  
The unit of the electric current density $j=|\bm j|$ is 
$2eJ/(pa^2)$ and is typically $\sim1.0\times 10^{13}$ A/m$^2$ 
for the polarization of magnet $p=0.2$ and the lattice constant $a=5$\AA.

\vskip0.5cm
\noindent
{\bf Acknowledgments}

\noindent
We thank  for Jan Masell and M. Ishida for useful discussions. 
This work was supported by JST CREST Grant Number JPMJCR1874, Japan, and JSPS KAKENHI Grant numbers 18H03676 and 26103006.

\end{document}